\documentclass[12pt,tightenlines,onecolumn,showpacs,showkeys,amsmath,amssymb]{revtex4}

\usepackage[paperwidth=210mm,paperheight=297mm,hmargin=4cm,vmargin=2.5cm]{geometry}
\usepackage[english]{babel}
\usepackage{graphicx}
\usepackage{subfigure}
\usepackage{dcolumn}
\usepackage{bm}
\usepackage{psfrag}
\usepackage{color}
\usepackage{fancyhdr}
\makeatletter

\begin{document}


\title{Excitation spectrum in two-dimensional superfluid $^4$He}

\author{F. Arrigoni}
 \affiliation{Dipartimento di Fisica, Universit\`a degli Studi
              di Milano, via Celoria 16, 20133 Milano, Italy}
\author{E. Vitali}
 \affiliation{Dipartimento di Fisica, Universit\`a degli Studi
              di Milano, via Celoria 16, 20133 Milano, Italy}
\author{D. E. Galli}
 \affiliation{Dipartimento di Fisica, Universit\`a degli Studi
              di Milano, via Celoria 16, 20133 Milano, Italy}
\author{L. Reatto}
\affiliation{Dipartimento di Fisica, Universit\`a degli Studi
              di Milano, via Celoria 16, 20133 Milano, Italy}
\date{\today}

\let\runauthor\@author
\let\runtitle\@title
\makeatother
\lhead{\runauthor}
\rhead{\runtitle}

\begin{abstract}
In this work we perform an {\it{ab-initio}} study of an ideal two-dimensional sample of $^4$He
atoms, a model for $^4$He films adsorbed on several kinds of substrates. Starting
from a realistic hamiltonian
we face the microscopic study of the 
excitation phonon-roton spectrum of the system at zero temperature. 
Our approach relies on Path Integral Ground State Monte Carlo projection methods, 
allowing to evaluate exactly the dynamical density correlation functions in imaginary time, and this
gives access 
to the dynamical structure factor of the 
system $S(q,\omega)$, containing information about the excitation spectrum $E(q)$, 
resulting in sharp peaks in  $S(q,\omega)$. The actual evaluation of $S(q,\omega)$ 
requires the inversion of the Laplace transform in 
ill-posed conditions, which we face via the Genetic Inversion via Falsification of Theories 
technique. We explore the full density range from the region of spinodal decomposition
to the freezing density, i.e. 0.0321 \AA$^{-2}$ -- 0.0658 \AA$^{-2}$.
In particular we follow the density dependence of the excitation spectrum,
focusing on the low wave--vector behavior of $E(q)$,
the roton dispersion, the strength of single quasi--particle peak, $Z(q)$, and 
the static density response function, $\chi(q)$.
As the density increases, the dispersion $E(q)$ at low wave--vector changes from a super--linear
(anomalous dispersion) trend to a sub--linear (normal dispersion) one, anticipating the 
crystallization of the system;
at the same time the maxon-roton structure, which is barely visible at low density,
becomes well developed at high densities and the roton wave vector has a strong density
dependence. Connection is made with recent inelastic neutron scattering results
from highly ordered silica nanopores partially filled with $^4$He.
\end{abstract}
 
\pacs{67.25.bh, 67.25.dt}

\keywords{superfluidity, two dimensional quantum fluids, elementary excitations, roton}

\maketitle
\section{Introduction}
Helium exists in two stable isotopes, $^4$He and $^3$He, which differ for their nuclear spin: 
$^4$He atoms are bosons with nuclear spin $I = 0$, while $^3$He atoms are fermions 
with nuclear spin $I = 1/2$.
The effective interaction among helium atoms is well described by a hard core potential plus an attraction 
arising from zero--point fluctuations in the charge distribution. 
The interaction results in a simple Lennard-Jones-like two-body spherically symmetric potential $v(r)$, 
for which accurate analytical expressions are known \cite{az79}.
The hamiltonian of the bulk system reads:
\begin{equation}
\label{hamiltonian}
\hat{H} = - \frac{\hbar^2}{2m}\sum_{i=1}^{N} \nabla_{i}^2
+ \sum_{i<j=1}^{N}v\left(|\vec{\hat{r}}_i - \vec{\hat{r}}_j|\right) \quad .
\end{equation}
where $m$ is the mass of $^4$He atoms.
Despite its very simple structure, helium exhibits numerous exotic phenomena in condensed form, whose theoretical 
explanation, in some aspects, is still a big challenge nowadays.
Along with the many fascinating physical features related
to the well known phenomenon of superfluidity \cite{legg}, which have been the object of several
theoretical and experimental efforts, a unique fingerprint of such a system is the
{\it spectrum} $E(q)$ of the elementary excitations. 

Excitations in $^4$He bulk systems have been 
extensively investigated after Landau's original conjecture \cite{land}  about the phonon-roton dispersion relation $E(q)$
and its connection with the definition of superfluidity in terms of a critical velocity.
In $1953$ Feynman showed that the shape of the phonon-roton spectrum can be justified on a quantum 
mechanical basis, relying on Bose statistics together with hard-core interactions \cite{fey1}.
Moreover, he suggested that the excitation spectrum of superfluid $^4$He may be investigated by
inelastic neutron scattering experiments. This was realized only almost
one decade later \cite{cowo}, beautifully confirming the original Landau's guess.
Actually, within the first Born approximation, the differential
cross section in a thermal neutron scattering experiment on a sample of $^4$He atoms,
a part from kinematical factors, is provided by the {\it dynamical structure factor}:
\begin{equation}
\label{sqw}
S(q,\omega) =  \frac{1}{2\pi N } \int_{-\infty}^{+\infty}dt\,e^{i\omega t} 
\langle e^{i\frac{t}{\hbar} \hat{H}}\,\hat{\rho}_{\vec{q}}\, 
e^{-i\frac{t}{\hbar} \hat{H}}\,\hat{\rho}_{-\vec{q}} \rangle
\end{equation}
where  the brakets indicate a ground state or a thermal average, $\hat{H}$ is the 
Hamiltonian of the helium system \eqref{hamiltonian}, and
$\hat{\rho}_{\vec{q}} = \sum_{i=1}^{N}\, \exp(-i \vec{q} \cdot \vec{\hat{r}}_i)
$, $\vec{\hat{r}}_i$ being the position operator of the $i$-th $^4$He atom,
 is the local density operator in Fourier space.
Sharp peaks in $\omega$ of $S(q,\omega)$ provide the spectrum of the
elementary excitations of the system.

On the theoretical side, a systematic effort has been devoted to pursue an accurate
description of the elementary excitations of the system. 
The original idea of Feynman-Cohen \cite{fey2} of introducing back--flow correlations to
improve variational excited states wave functions
(later on extended by the correlated--basis--functions strategy \cite{cbf}),
has flown into the excited states Shadow wave functions technique \cite{eswf} (SWF);
SWF reproduced the experimental bulk dispersion relation $E(q)$
up to a high accuracy level \cite{swf2} and even confirmed \cite{swf2,roto} the physical picture of a roton 
as a microscopic smoke ring \cite{feyl}.
A further turn in the study of excited states of superfluid $^4$He was given by
the advent of exact simulation methods for interacting Bose particles. It is
not yet possible to perform a direct exact computation of excited states due
to the sign problem. However, it is possible to extract dynamical properties from
exact correlation functions in imaginary time \cite{gube}.
This has been worked out by Path Integral Monte Carlo (PIMC) at finite
temperature \cite{cepe} or by Ground State Monte Carlo \cite{rept,gift} at $T = 0$ K.
Indeed, such functions contain information on excited states of the system. In particular
the density correlation function is related to $S(q,\omega)$ by an inverse 
Laplace transform. Due to discretization and statistical noise, the 
mathematical problem of extracting $S(q,\omega)$ is ill-posed, but powerful
inversion methods have been introduced recently \cite{asm,gift}
and reliable results on the excitation spectrum of superfluid $^4$He
have been obtained \cite{gift,gift2,gift3}.

Bosons in two dimensions (2D) are of great theoretical interest
because the standard scenario of superfluidity associated with Bose-Einstein
condensation (BEC) is not appropriate. In fact, in $2D$ and in almost
$2D$ systems the order parameter, i.e. the condensate wave function, $\psi(\vec{r})$,
vanishes at any finite temperature for a bulk system. The notion of
long range order is replaced by that of topological long range order \cite{ktb}
with correlation function of the local order parameter decaying algebrically very slowly
to zero. Notwithstanding a vanishing order parameter, a superfluid response
is theoretically predicted up to a temperature where vortex and
antivortex pairs unbind. These predictions have been beautifully
confirmed by experiments \cite{reppy}. 
Therefore a 2D Helium system is an interesting
microscopic model for quasi-two dimensional 
many--body quantum systems \cite{gfmc,exp2}: helium films on suitable 
substrates.
For most substrates the interaction potential between the helium atoms
and the substrate is much stronger (as it is the case of He-graphite interaction)
than the He--He interaction and the
helium atoms are adsorbed in a well-defined layer structure.
Typically, only the first or the first two layers are strongly 
influenced by the details of the
helium--substrate interaction.
Several different physical realizations of substrates have been
investigated, both in experimental and in theoretical works.
For many substrates the closest He atoms to the substrate are
disordered and localized, they form what the experimentalists call a ``dead
layer''. Beyond that the first layer of mobile atoms are superfluid
and can be well represented by a strictly 2D model. The experimental
study of $S(q,\omega)$ of this film has shown the existence of elementary
excitations with a phonon-maxon-roton structure \cite{godfrin}. 
A favorite substrate
for adsorption studies is graphite because it offers rather extended
regions of perfectly flat basal planes. At first sight this
might be considered as an ideal situation for using the 2D model.
This is not so for the first adsorbed layer because the adsorption potential
is strongly corrugated. The consequence of the corrugation is that at
low temperature the $^4$He atoms form an ordered structure, either
a triangular lattice that is commensurate with the substrate or, at higher coverage,
an incommensurate triangular solid \cite{cole}.
Experimentally no evidence has been found for superfluidity in
the first adsorbed layer on graphite. Superfluidity has been found
only in additional layers for which the 2D model can be used
as a reasonable approximation.

Computation of the spectrum of elementary excitations of $^4$He is of
interest on one hand to uncover the dependence of rotons on the
dimensionality of the system. On the other hand this theoretical
input is useful for the interpretation of scattering experiments from
adsorbed $^4$He. 

Excitations for 2D $^4$He have been studied by correlated
basis function theory \cite{cleme}.
As far as we know, the only existing {\it ab initio}
quantum Monte Carlo (QMC)
 calculation of excitations in 2D $^4$He
has been performed with variational theory using shadow wave functions \cite{gris}. 
As mentioned above exact QMC techniques are able to
give access to estimations of $S(q,\omega)$ via exact calculations of dynamical correlation
functions in imaginary time.
The Path Integral Ground State (PIGS) method \cite{pigs} and in particular the Shadow Path Integral Ground State (SPIGS)
method \cite{spigs1,spigs2} together with the Genetic Inversion
via Falsification of Theories (GIFT) method \cite{gift}
have been applied to bulk $^4$He systems \cite{gift,gift2},
to adsorbed $^4$He systems \cite{gift3} and even to a pure 2D $^3$He system \cite{gift4}, (via a quite sophisticated novel strategy).
Here we apply such approaches to address the calculation of dynamical properties of a pure 2D $^4$He system
at zero temperature.

The article is structured as follows: in the next Section we sketch the methodology; 
in Section III we present and discuss the results and our conclusions are in Section IV.
In the Appendix we give the results of a variational computation of the ground state
properties of $^4$He in 2D based on Shadow Wave Functions (SWF) that are a byproduct
of the exact SPIGS computation of Sect. III.   

\section{Methodology}
We focus thus on a strictly $2$D collection of $N$ structureless spinless bosons at zero temperature.
The hamiltonian operator is \eqref{hamiltonian}. 
We let $\psi_0(\mathcal{R})$ be the Ground State of $\hat{H}$, where we use the notation 
$\mathcal{R} = (\vec{r}_1, \dots, \vec{r}_N)$. The basic relation underlying QMC projection methods is the following:
\begin{equation}
\label{projection}
\psi_0(\mathcal{R}) \propto \lim_{\lambda \to +\infty} e^{-\lambda  \hat{H} } \, \psi_T(\mathcal{R}) 
\end{equation}
where $\psi_T(\mathcal{R})$ is any many-body wave function with non zero overlap on 
$\psi_0(\mathcal{R})$. The operator $e^{-\lambda  \hat{H} }$ can be seen as the
evolution operator, $e^{-i(t/\hbar) \hat{H} }$, written for imaginary times;
in this way, $\psi_0(\mathcal{R})$ turns out to be
the limit of the imaginary time evolution of $\psi_T(\mathcal{R})$, with
$\lambda$ playing the role of imaginary time.
A Trotter decomposition:
\begin{equation}
e^{-\lambda \hat{H}} = \left(e^{-\delta \tau  \hat{H} } \right)^M, \quad \delta \tau  = \frac{\lambda}{M}
\end{equation}
together with an (analytical or numerical) approximation for the imaginary time propagator:
\begin{equation}
\langle \mathcal{R} | e^{-\delta \tau \hat{H} } | \mathcal{R}'\rangle= \mathcal{G}(\mathcal{R},\mathcal{R}',\delta\tau) + \mathcal{O}(\delta\tau^m) \quad ,
\end{equation}
where the order $m$ depends on the approximation, allows to build up an approximate expression for 
the ground state wave function of the form:

\begin{equation}
\label{pigs}
\psi_0(\mathcal{R}) \simeq 
 \int d\{\mathcal{R}_i\} \,
\mathcal{G}(\mathcal{R},\mathcal{R}_1,\delta\tau) \dots \mathcal{G}(\mathcal{R}_{M-1},\mathcal{R}_M,\delta\tau)
\, \psi_T(\mathcal{R}_M)
\end{equation}
where we have omitted an overall normalization factor.
Any expectation value of an operator diagonal in coordinate representation (or of the Hamiltonian operator):

\begin{equation}
\label{static}
\langle \psi_0 | \, \hat{O} \, | \psi_0 \rangle
\end{equation}
is expressed as a multidimensional average of a function $O(\mathcal{R})$ over a probability density of the form:

\begin{equation}
p(\{\mathcal{R}_i\} ) = \frac{1}{\mathcal{Z}} \, \psi_T(\mathcal{R}_1) \prod_{i=1}^{2M}
\mathcal{G}(\mathcal{R}_{i},\mathcal{R}_{i+1},\delta\tau)\, \psi_T(\mathcal{R}_{2M}) 
\end{equation}
which can be sampled using Metropolis algorithm.
The results can be considered {\it exact} in 
the sense that the errors arising from approximations can be reduced under the level of the 
statistical noise via a suitable choice of the time step $\delta\tau$ and the total projection time $\lambda = M\delta \tau$.
Of course this also assumes that the results, for large enough $\lambda$, are independent on the choice of $\psi_T$.
This has been verified \cite{patate}, even by starting with $\psi_T$ of a liquid for the solid phase or of a solid
for a liquid phase one finds convergence to the correct result. Notwithstanding this, a judicious
choice of $\psi_T$ is important to accelerate convergence of $\psi_T$ to $\psi_0$, i.e. a smaller value
of $\lambda$ is needed, and to reduce the variance of the results.
What has been described here is the PIGS method, or the SPIGS method if $\psi_T$ is a SWF.

This calculation scheme can be straightforwardly generalized to evaluate dynamical imaginary time correlation functions:

\begin{equation}
\label{dynamic}
\langle \psi_0 | e^{\tau \hat{H}} \, \hat{O} \, e^{-\tau \hat{H}} \hat{O}^{\dagger} | \psi_0 \rangle \quad .
\end{equation}
The particular choice:

\begin{equation}
\label{fqt}
F(\vec{q},\tau) = \langle \psi_0 | e^{\tau \hat{H}} \, \hat{\rho}_{\vec{q}} \, e^{-\tau \hat{H}} \, \hat{\rho}_{-\vec{q}} \, | \psi_0 \rangle
\end{equation}
provide the intermediate scattering function in imaginary time, which is related to the dynamical structure factor by the relation:

\begin{equation}
\label{ip}
F(q,\tau) = \int_{0}^{+\infty}d\omega \, e^{-\tau \omega} S(q,\omega)
\end{equation}
Thus, the estimation of $S(q,\omega)$ requires to invert the integral relation \eqref{ip} in ill-posed
conditions, since $F(q,\tau)$ is known only on a discrete and finite set of instants $\tau$ (typically $\tau = n \delta\tau$, $n = 0, \dots, \overline{n}$) and is affected
by a statistical uncertainty arising from the stochastic Monte Carlo calculation.
Despite the well known difficulties related to the inversion of the Laplace
transform in ill-posed conditions, the evaluation of $S(q,\omega)$ starting
from the QMC estimation of $F(q,\tau)$ \eqref{fqt} has been proved to 
be fruitful for several bosonic systems using a recent technique called Genetic Inversion via Falsification
of Theories (GIFT)\cite{gift}.
GIFT is a statistical inversion method: it samples a suitable space of spectral functions looking for models
compatible with the QMC data $F(q,\tau)$ via a stochastic search scheme relying on genetic algorithms.

\section{Simulation details and Results}
In our simulations of $^4$He in 2D we have used as interatomic potential
$v(r)$ the $1979$ Aziz potential \cite{az79} and $N = 120$ number
of atoms with periodic boundary conditions.
As propagator $\mathcal{G}(\mathcal{R},\mathcal{R}',\delta\tau)$
we have used the pair--product approximation \cite{pair} with $\delta\tau = 1/160$ K$^{-1}$,
a value that we have verified to be small enough for the adopted propagator.
As projection time $\lambda$ we have used $\lambda=1.1$ K$^{-1}$ and typical length of the
simulation is $3\times 10^6$ Monte Carlo steps (MCS);
$F(q,\tau)$ has been computed over the range $1 \div 90\times \delta\tau$.
A typical run starts from a triangular lattice configuration which quickly ``melts'',
when the density is not too large,
in few thousand MCS leading to disordered configurations allowing to simulate the
liquid phase without memory of the starting point.
When the density is large enough the system remains in an ordered state
as shown by the presence in the static structure factor $S(q)$ of sharp
Bragg peaks corresponding to triangular solid $^4$He.
Only in the density range of the liquid--solid transition one
gets convergence to two different states depending on the initial
configuration: starting from a disordered configuration the system
remains disordered whereas it remains ordered when started from
the ordered configuration. The energies of the two states
take different values, the lowest one represents the stable
phase and the higher energy one represents a metastable state
for a single phase state.

\begin{figure}[t!]
\includegraphics*[scale=0.45]{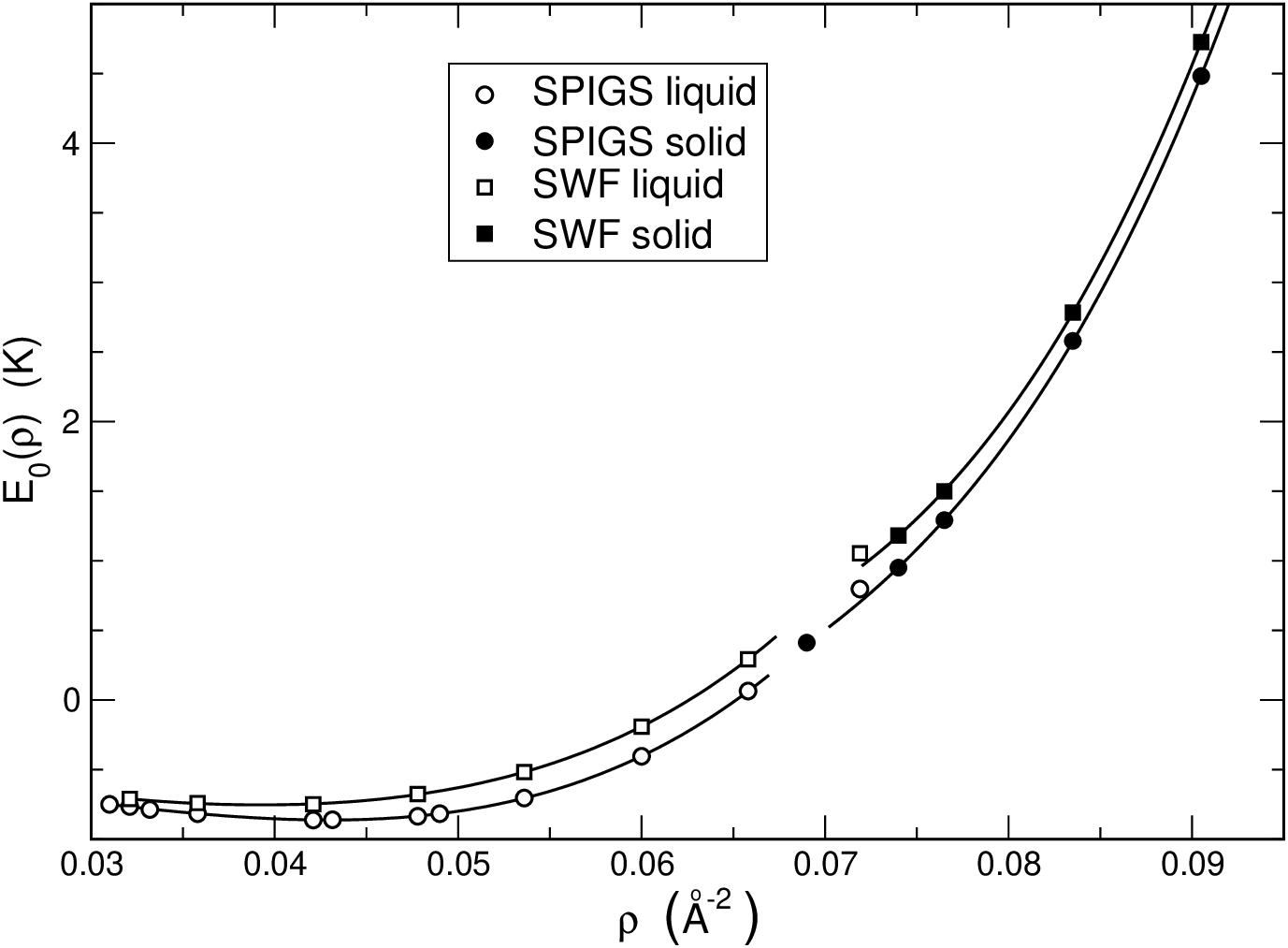}
\caption{\label{fig:eos} (squares) variational energies in the liquid (empty symbols) and in the
solid (filled symbols) phase. (circles) Exact energies in the liquid (empty symbols) and in the
solid (filled symbols) phase. The curves are the interpolated equations of state, and are truncated in the coexistence region.}
\end{figure}
The average value of the hamiltonian operator as a function of the density provides the 
equation of state of the system. We have computed the energy in a quite 
large range of densities both in the liquid and in the solid phase and the results are shown in 
Fig.~\ref{fig:eos} and listed in tables \ref{table:spigsliq} and 
\ref{table:spigssol} of the Appendix. The SPIGS results agree within
the statistical uncertainty with the result of a Green Function Monte Carlo
(GFMC) computation \cite{gfmc}. One can also see that the SWF variational
results follow quite closely the exact SPIGS results both in the liquid
and in the solid phase, thus confirming the accuracy of the SWF as in the 3D case.
In order to determine the melting and freezing densities, the energies have
been fitted with a third degree polynomial in density in the solid phase,
and a fourth degree polynomial in the liquid phase.
We write the fitting function as
\begin{equation}
\label{eosl}
E_l(\rho)=E_0 + A \left (\dfrac{\rho - \rho_0}{\rho_0} \right )^2 + B \left (\dfrac{\rho - \rho_0}{\rho_0} \right )^3
+ C \left (\dfrac{\rho - \rho_0}{\rho_0} \right )^4
\end{equation}
in the liquid phase, where a minimum energy $E_0$ at the equilibrium density $\rho_0$ is present.
The last term, which is not typical in literature, turned out to be necessary in order to obtain a
good fit in the whole density range here considered.
On the other hand, in the solid phase we use the expression:
\begin{equation}
\label{eoss}
E_s(\rho)=\alpha + \beta \rho + \gamma \rho^2 + \delta \rho^3 \quad .
\end{equation}
The obtained fitting parameters, together with their statistical uncertainties, are
listed in table \ref{table:fitex} of the appendix.
The interpolation curves, depicted in Fig.~\ref{fig:eos}, are truncated in the
coexistence region, delimited by the melting and freezing densities $\rho_m$ and $\rho_f$.
$\rho_m$ and $\rho_f$ have been estimated using the Maxwell construction, and they are given in Table 
\ref{table:fitex} of the Appendix.

In Fig.~\ref{fig:sound} we show some quantities like the pressure $p$, the chemical potential $\mu$,
the compressibility $\kappa$ and the sound velocity $v_s$ in the liquid and in the solid phase;
such quantities have been obtained from $E_0(\rho)$
via the expressions:
$p(\rho)=\rho^2 \frac{\partial E_0(\rho)}{\partial \rho}$,
$\mu(\rho)=E_0(\rho)+p(\rho)/\rho$,
$\kappa(\rho)=(\rho \frac{\partial p(\rho)}{\partial \rho})^{-1}$
and $v_s(\rho) = \sqrt{\frac{1}{m} \frac{\partial}{\partial \rho} \left(\rho^2 \frac{\partial E_0(\rho)}{\partial \rho}\right)}$. In the solid phase $v_s$ represents the velocity of
the longitudinal sound mode.
\begin{figure}[t!]
\includegraphics*[scale=0.45]{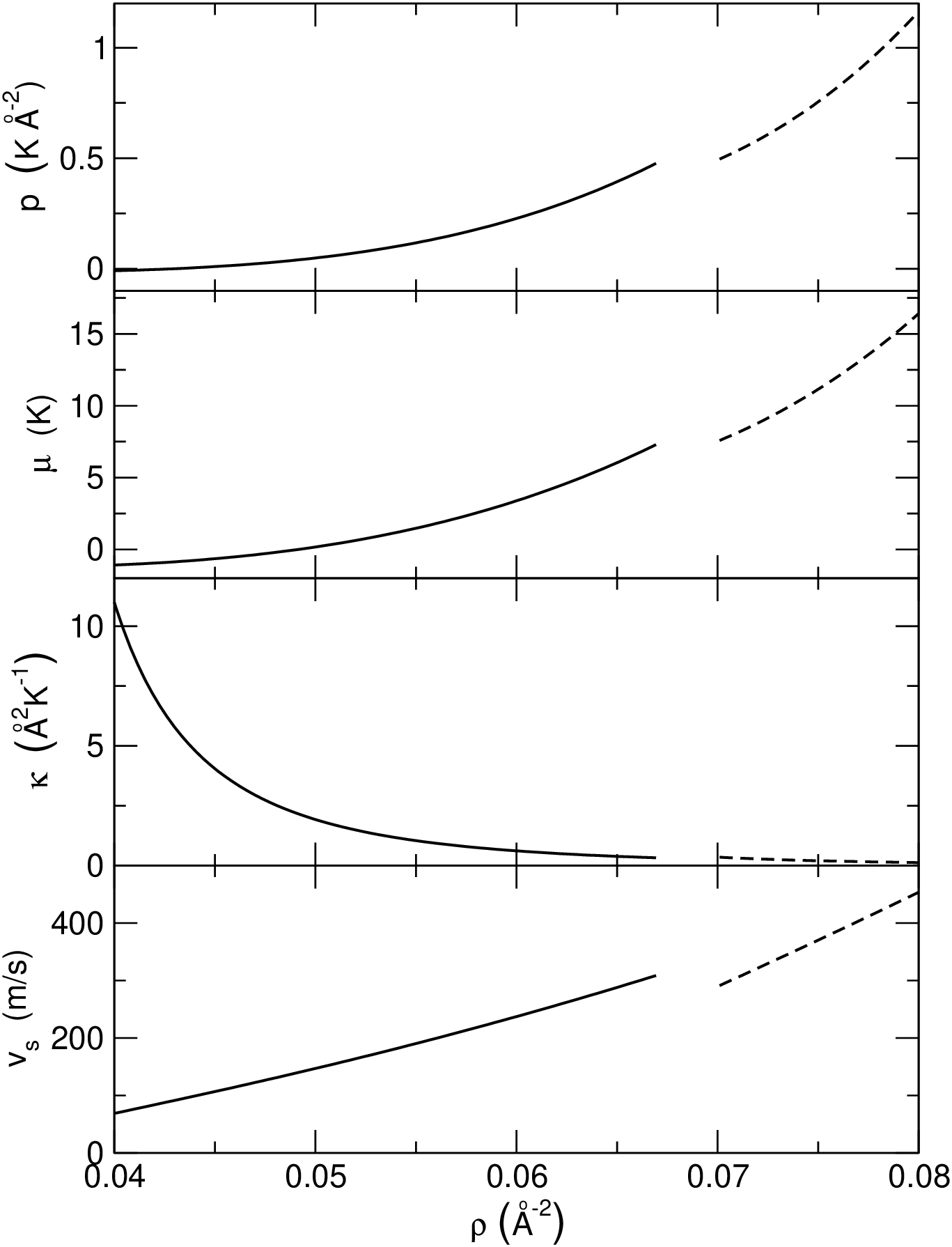}
\caption{\label{fig:sound} Thermodynamical properties derived from the equation of state as functions of the density;
(solid line) liquid phase; (dashed line) solid phase:
(top panel) pressure $p$; (middle upper panel) chemical potential $\mu$; (middel lower panel) compressibility $\kappa$;
(bottom panel) sound velocity $v_s$.}
\end{figure}

\begin{figure}[t!]
\includegraphics*[scale=0.45]{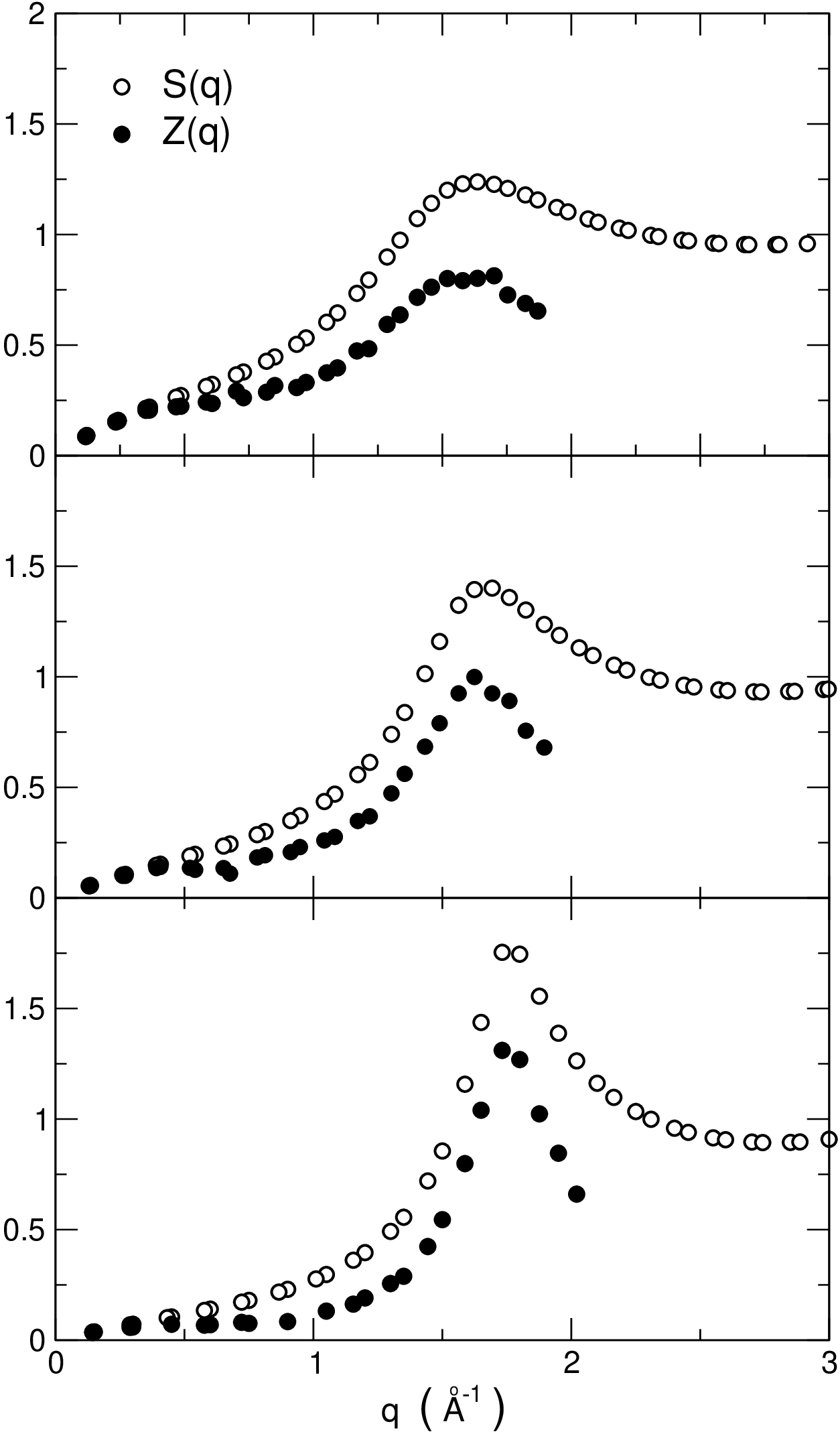}
\caption{\label{fig:sq} SPIGS estimations of the static structure factor $S(q)$ (empty circles) and strenght
of the single particle peak $Z(q)$ (filled circles) at the densities $0.04315$\AA$^{-2}$ (upper panel),
$0.0536$\AA$^{-2}$ (middle panel)
and $0.0658$\AA$^{-2}$ (lower panel).}
\end{figure}
In Fig.~\ref{fig:sq} we show the SPIGS result for the static structure factor 
$S(q) = \langle \psi_0 | \, \hat{\rho}_{\vec{q}} \, \hat{\rho}_{-\vec{q}} \,| \psi_0 \rangle$
for a density close to the equilibrium one and at a density close to freezing. 
It is evident the emergence of 
more structure as the density increases towards the freezing density. Moreover, we emphasize the 
linear behavior of $S(q)$ for $q \to 0$ which manifests itself at very small wavevectors.
This is due to the zero--point motion of long wavelength phonons \cite{rech}.
 
We have computed the dynamical correlation functions for imaginary time
$F(q,\tau)$ at six densities, and namely $0.0321$\AA$^{-2}$, $0.0421$\AA$^{-2}$,
$0.04315$\AA$^{-2}$, $0.049$\AA$^{-2}$, $0.0536$\AA$^{-2}$ and $0.0658$\AA$^{-2}$ in the liquid phase.
From $F(q,\tau)$, the GIFT method allows to reconstruct the dynamical structure 
factor of the sample, $S(q,\omega)$. 
An example of $F(q,\tau)$ and of the reconstructed $S(q,\omega)$ is shown in Fig.~\ref{fig:excspec}.
\begin{figure}[t!]
\includegraphics*[scale=0.45]{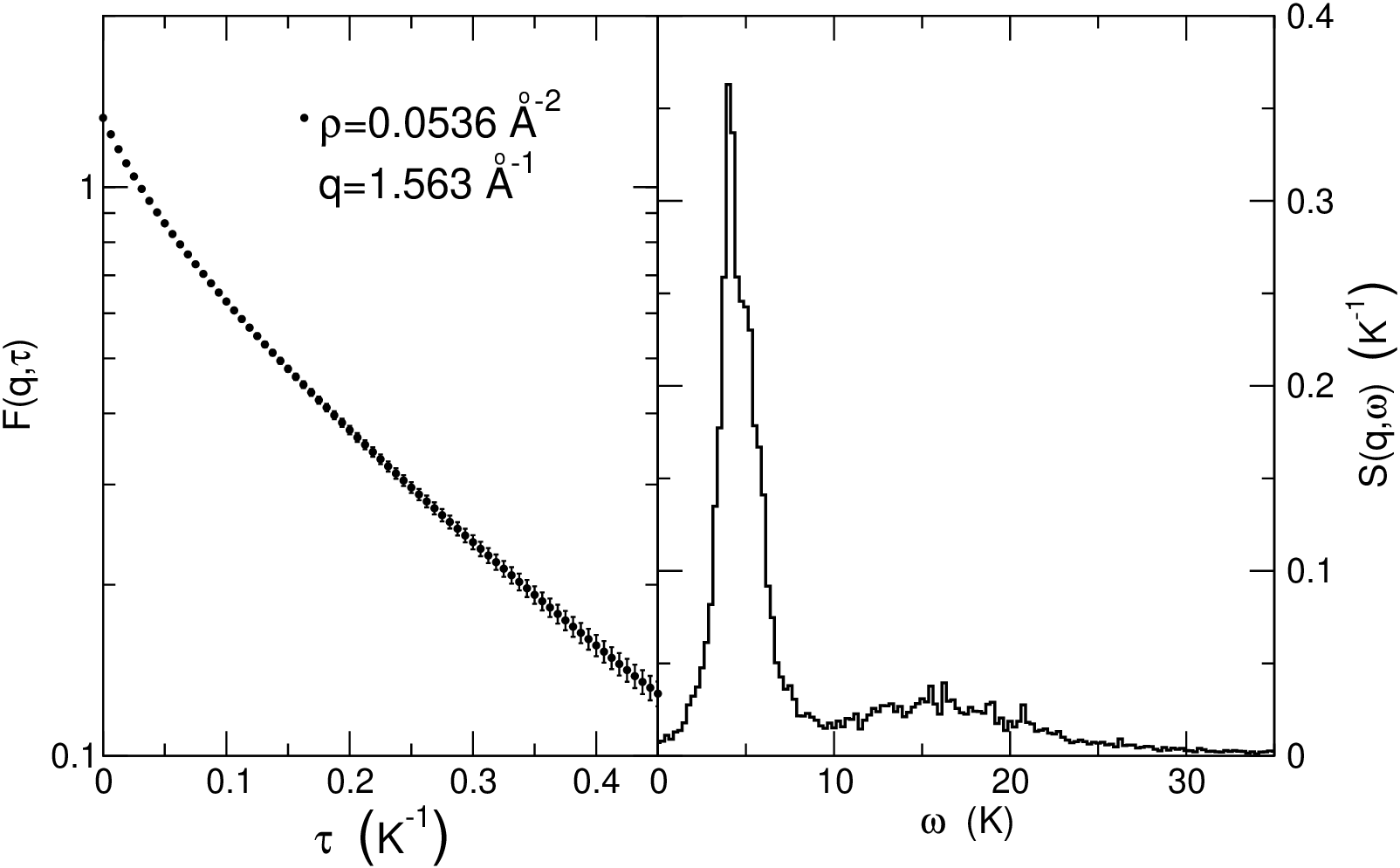}
\caption{\label{fig:excspec} (left panel) An example of QMC evaluation of an 
imaginary time correlation function $F(q,\tau)$, defined in \eqref{fqt}.
We have plotted the $\tau$-dependence of $F(q,\tau)$ for a given wave-vector
$q$ (see the legend) in logarithmic scale to show the asymptotic single
exponential behavior governed by the elementary excitation energy. (right panel)
Reconstructed $S(q,\omega)$: one can see the sharp elementary excitation peak 
together with the higher energy broad multiphonon contribution.   }
\end{figure}
$S(q,\omega)$ in general has a sharp peak in $\omega$ and this defines the energy $E(q)$
of the excitation for the given wave vector $q$. In addition there is a
much broader peak at larger energy and this represents the so called
multiphonon contribution to $S(q,\omega)$.
The elementary excitation peak in the reconstructed $S(q,\omega)$ has a finite width.
This width can have two different origins. As discussed in Ref.~\cite{gift}, even if
the system has an infinitely long--lived excitation the peak in the reconstructed
$S(q,\omega)$ has a finite width because the inversion method can only identify the
excitation energy with a certain uncertainty due to the limited and noisy information
on $F(q,\tau)$. In this case the full width at half maximum (FWHM) can be taken as
a measure of statistical uncertainty of the excitation energy. Under certain circumstances
even at $T = 0$ K an elementary excitation acquires a finite lifetime when it can
decay into two or more excitations. This happens, for instance, for the maxon excitation
in superfluid $^4$He in 3D at large pressure when the maxon energy is larger than
twice the roton energy. In this case the excitation peak has an intrinsic finite 
linewidth and its FWHM is a measure of the inverse life-time of the excitation. 
Under such circumstances we expect that the width of the
reconstructed $S(q,\omega)$ has also a contribution of intrinsic origin due to such
physical processes, even if it is difficult to quantify precisely how large
this intrinsic contribution is from the overall FWHM.

The integral over all $\omega$ of $S(q,\omega)$ is equal to the static structure
factor $S(q)$. An important information is contained in the strength $Z(q)$
of the elementary excitation peak, i.e. the integral of $S(q,\omega)$ limited
to the main peak. The ratio $Z(q)/S(q)$ gives the probability
that in the scattering process there is emission of a single elementary
excitation whereas $1-Z(q)/S(q)$ gives the probability of emission of
other excitations, the so-called multiphonon processes. The behavior of $Z(q)$
is shown in Fig.~\ref{fig:sq} at three densities.

In Fig.~\ref{fig:excspec} we show the obtained dispersion relation $E(q)$ for four 
values of the density. The reported bar represents the FWHM of the main peak in $S(q,\omega)$.

\begin{figure}[t!]
\includegraphics*[scale=0.45]{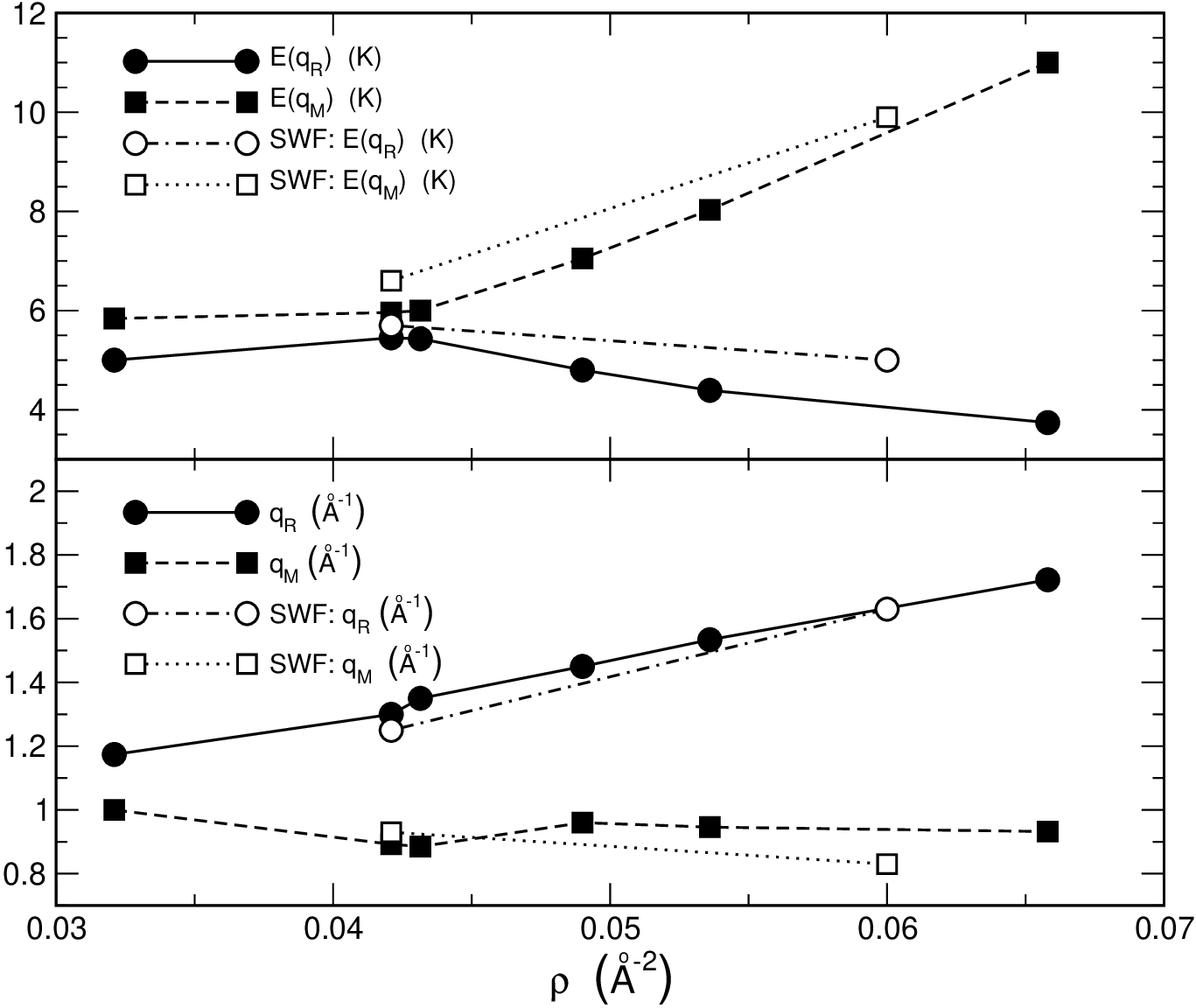}
\caption{\label{fig:dettrm} Density dependence of the wave-vector and of the energy of the maxon
($q_M$, $E(q_M)$) and the roton ($q_R$, $E(q_R)$) modes. Lines are guides to the eye.}
\end{figure}

At the lowest density, $0.0321$\AA$^{-2}$, which is near the spinodal decomposition, the 
excitation spectrum shows a large flat region, and a very weak roton minimum. At small 
wavevector the spectrum shows an anomalous dispersion, i.e. a positive curvature.
At the density $\rho_0=0.04315$\AA$^{-2}$ close to equilibrium 
the phonon-maxon-roton structure starts to be visible but maxon energy does
not differ by more than $10\%$ with respect to the roton energy.
As the density further increases the maxon-roton region becomes more
and more prominent until, 
\begin{figure}[t!]
\includegraphics*[scale=0.45]{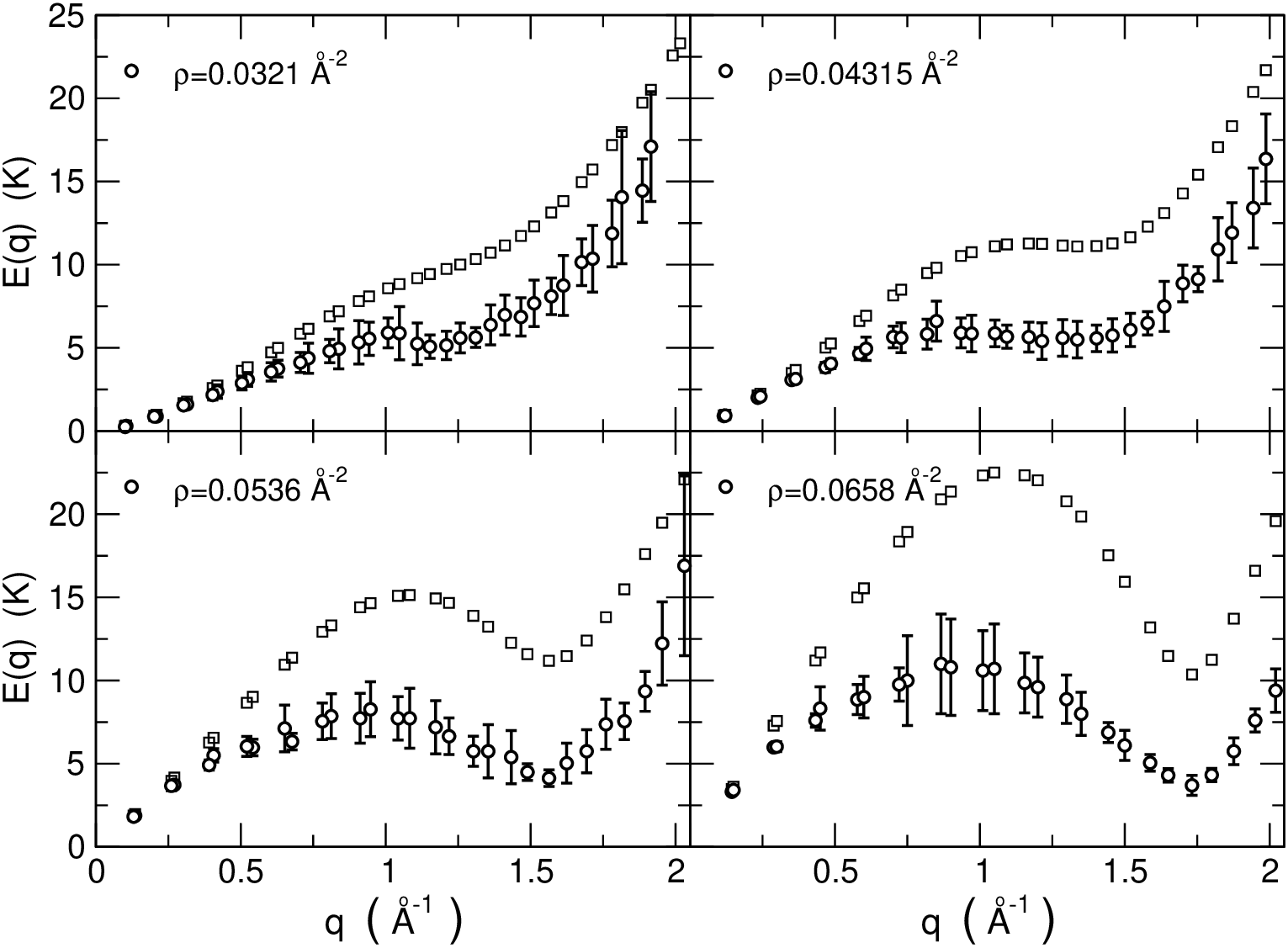}
\caption{\label{fig:excspec} (circles) Excitation spectrum form GIFT reconstructions of SPIGS evaluations of 
imaginary time correlation functions in the liquid phase, together with Feynman spectrum (squares), 
at four densities as shown in the legends.}
\end{figure}
at the highest density $0.0658$\AA$^{-2}$, near the freezing point, the 
maxon energy is about three times the roton energy.
At the larger density the peaks in the maxon region
are quite broadened, as it
is evident from the error bars in Fig.~\ref{fig:excspec}; we believe that
in this case the linewidth largely represents an intrinsic effect due to
the fact that a maxon can decay into two rotons because its energy is more
than twice the roton energy. This fact is known experimentally \cite{cowo2}
and theoretically \cite{swf2,roto} in 3D superfluid $^4$He at density in the
region of freezing. In Fig.~\ref{fig:dettrm} we plot the energy and the
wave vector of roton and of maxon as function of
density. It can be noticed that the roton energy in 2D
(from 5.5 to 3.8 K depending on density) is significantly
below the value in $3D$ (from 8.6 K at equilibrium to 7.2 
K at freezing density). It can also be noticed that the
roton wave-vector has a significant density dependence while the
maxon wave-vector is almost density independent. 

\begin{figure}[t!]
\includegraphics*[scale=0.45]{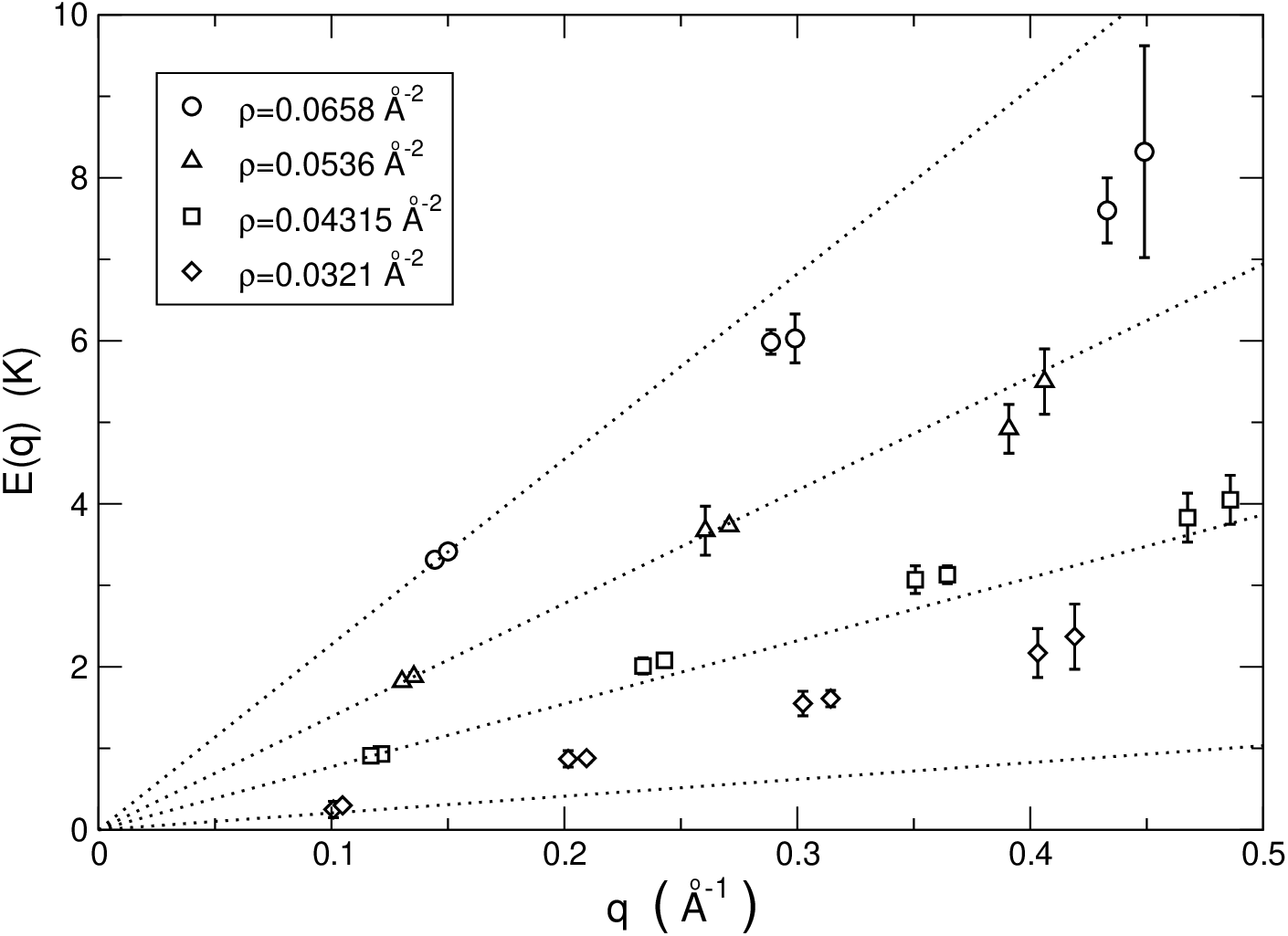}
\caption{\label{fig:anom} Small wave vectors behavior of the estimated dispersion relation $E(q)$
for the densities $0.0321$\AA$^{-2}$ (diamonds), $0.04315$\AA$^{-2}$ (squares) $0.0536$\AA$^{-2}$ (triangles)
and $0.0658$\AA$^{-2}$ (circles).
The dotted straight lines represent the linear behavior from which the $E(q)$ significantly deviate.
}
\end{figure}

In Fig.~\ref{fig:excspec} we show also the Feynman spectrum, $E_F(q) = \hbar^2 q^2/2mS(q)$, obtained 
using our estimation of $S(q)$. Feynman dispersion relation is accurate, as it is well known, only in 
the low wave vectors region. The discrepancy between $E(q)$ and $E_F(q)$ increases with the density: $E_F(q)$
is more than twice $E(q)$ near the freezing point.
We notice also that the present $E(q)$ is 
in good agreement with the variational result at the
equilibrium density obtained using Shadow Wave functions in Ref.~\onlinecite{eswf}.
At larger density the variational roton energy is about 1 K above the present
result. 
In Fig.~\ref{fig:anom} we show more details about the low $q$ behavior of the dispersion relation
at four considered densities.
It is apparent that the phononic dispersion is superlinear for the two lowest densities, and becomes
sublinear at larger densities up to the freezing point. 
This is qualitatively similar to what happens in superfluid $^4$He in 3D.

With respect to the strength of the quasi-particle peak $Z(q)$, 
at all densities $Z(q) \simeq S(q)$ at small $q$, i.e. the collective
excitation peak almost exhausts the f-sum rule and multiphonon contributions
are negligible. At equilibrium density the roton peak has about $2/3$ of the 
full integrated intensity and $1/3$ is due to multiphonon contribution.
This multiphonon contribution is larger than in 3D and we attribute
this to the fact that equilibrium density in 2D is rather low where short
range order is not very pronounced. Only near freezing the multiphonon
contribution of the roton is small (of order of $20\%$) as in 3D superfluid
$^4$He.

Studies of the elementary excitations of $^4$He in restricted geometry
have been performed by inelastic neutron scattering on $^4$He confined
in a number of nanopore materials. Of special relevance is a recent study \cite{pori} 
of $^4$He in smooth cylindrical silica pores of diameter of about 28 \AA.
When the pores are filled with $^4$He experiment shows the presence
of phonon-maxon-roton excitations with a dispersion relation very similar
to that of bulk $^4$He. Such excitations are interpreted as propagating in
the central part of the pore. An additional roton excitation at smaller
energy is present and this is interpreted as roton confined in a compressed
layer close to the cylinder wall.  When the pores are only partially
filled with $^4$He the compressed layer rotons are still present,
whereas the bulk--like phonon-maxon-roton branch disappears. In its place
there is a modified phonon-maxon-roton branch with a decreased energy of the
maxon ($11$ K instead of 14 K in the bulk) and a roton energy only 2 K
below the maxon (the energy difference beyween maxon and roton in bulk $^4$He
is about 5 K at equilibrium density). 
In addition this new roton is found at a shifted wave-vector, at 1.78
\AA$^{-1}$ in place of 1.92 \AA$^{-1}$ of the bulk one. This modified
maxon-roton branch has been interpreted as propagating in a thin film
inside the unfilled pore and connection has been made with the excitations
in 2D $^4$He as computed in Ref.~\onlinecite{gris}. Indeed some similarity
between the dilute layer modes of experiment and the present results
for $^4$He in 2D is present, such as the reduced energy difference between
maxon and roton and a reduced wave vector $q_R$. On the other hand some
significant difference is present. For instance we find $q_R \simeq 1.75$
\AA$^{-1}$ at a density close to freezing but here the roton energy is
about 4 K, less than half the value of the dilute layer mode.
Of course there is a difference between the present mathematical
2D system and the finite curvature of the $^4$He film in an unfilled
pore of the experiment. It is unclear if this might be the origin of
that difference for the roton energy.

\begin{figure}[t!]
\includegraphics*[scale=0.45]{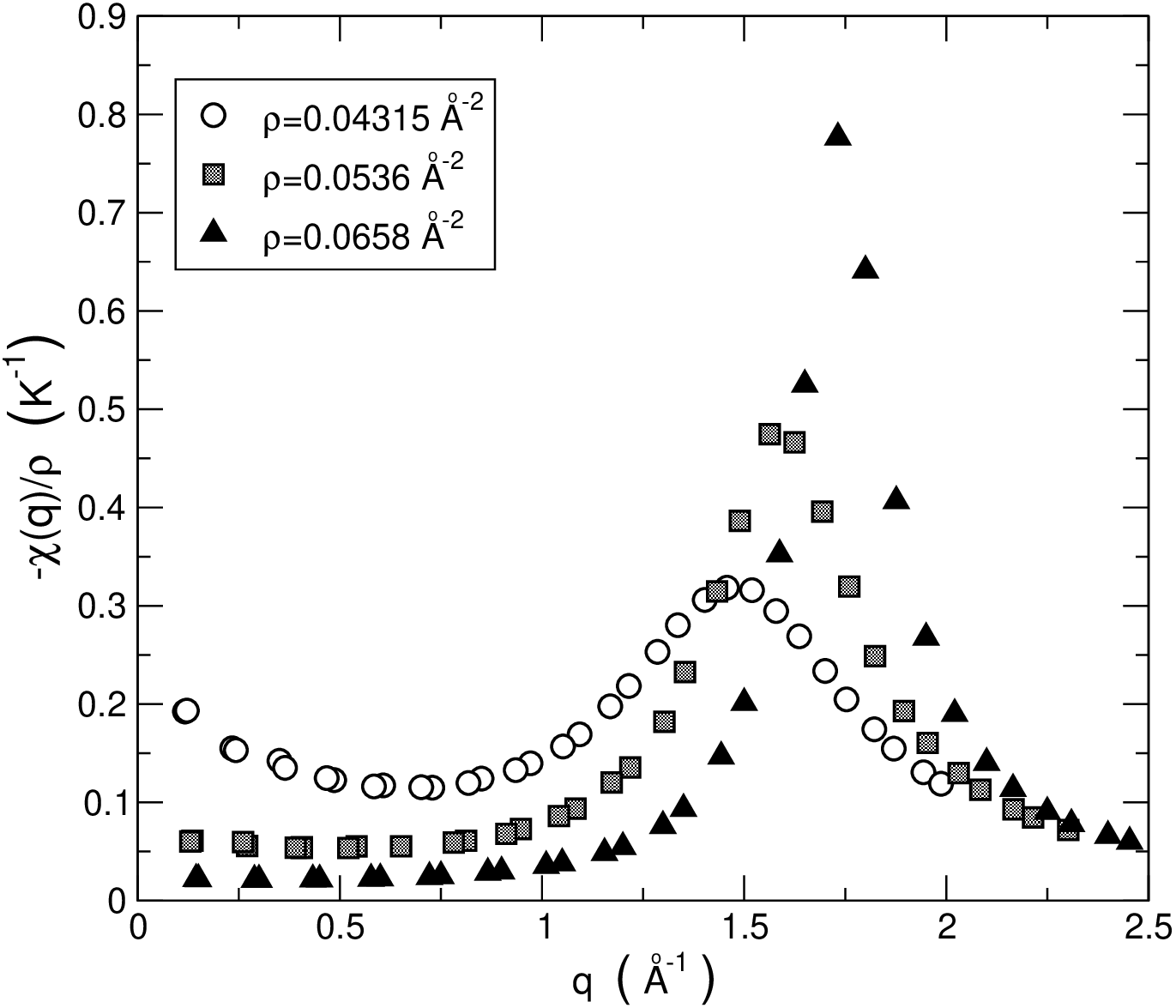}
\caption{\label{fig:chiq} Density response function extracted from the dynamical structure factor
at the densities $0.04315$\AA$^{-2}$ (circles), $0.0536$\AA$^{-2}$ (squares) and $0.0658$\AA$^{-2}$ (triangles).}
\end{figure}

Finally, we obtained from the $-1$--moment of $S(q,\omega)$ also the static density
response function, $\chi(q)$, which is shown in Fig.~\ref{fig:chiq}.
As in 3D $\chi(q)$ is dominated by a peak at the roton wave-vector.
One can notice that at the equilibrium density $\chi(q)$ has an
enhancement at small $q$ which is absent in 3D \cite{gift}.
This is another manifestation that the ground state of $^4$He
in 2D is at low density where atoms are not strongly coupled as
in 3D.

\section{Conclusions}
We have presented the first {\it{ab initio}} QMC computation of the
excitation spectrum of superfluid $^4$He in 2D starting from exact
density correlation functions in imaginary time and using advanced
inversion methods to infer the dynamical structure factor $S(q,\omega)$.
We find well defined excitations in the full density range
of the superfluid but significant differences are present with 
respect to $^4$He in 3D. In 3D the excitation spectrum over
the full density range from the equilibirum density to the freezing one has a well defined
phonon-maxon-roton structure with the maxon energy $E_M$ larger
by at least 50\% than the roton energy $\Delta_R$.
In 2D the excitation spectrum evolves with density
from maxon and roton almost coaleshing in a plateau 
at density close to the spinodal to a well defined maxon-roton
structure at density above the equilibrium one with $E_M/\Delta_R$ 
becoming as large as 3 at freezing. At the same time the wave-vector
$q_R$ of the roton has a strong density dependence whereas 
that of the maxon is almost density independent. This strong
evolution with density of the shape of the excitation spectrum
is due to the large density range of existence of the fluid in 2D,
the freezing density is more than twice the spinodal density while
in 3D the freezing density is only 60\% larger than the spinodal
one. At the 2D equilibrium density the maxon-roton structure is rather
weak with the maxon energy only $10\%$ larger than the roton energy.
This is due to the low value of the equilibrium density
so that the amount of short range order is rather small. At the
same time in the phonon region there is a strong anomalous dispersion
(i.e. $E(q)$ has a positive curvature). As a consequence of the
shape of $E(q)$, over an extended region of $q$ and of density,
the elementary excitations are expected to have a finite lifetime
even at $T=0$ K, because they can decay into other excitations. We
find evidence for this finite lifetime from our computation but the
present method does not allow to quantify this. 

It has been suggested \cite{pori} that the excitation spectrum of 2D
$^4$He might be relevant for the interpretation of the excitations
of $^4$He partially filling smooth cylindrical silica pores as measured
by inelastic neutron scattering. We indeed find some similarity
between our results and the experimental ones. However we find a
strong disagreement in the value of the roton energy which is well
beyond the uncertainty of the present theory. This discrepancy might be
due to a curvature effect that is present in the pore but not
in the present computation. It will be interesting to extend the 
present computation to the case of a pore geometry; present 
developments of QMC techniques are such that this is a feasible
project.

\section{Acknowledgements}
This work has been supported by Regione Lombardia and CINECA Consortium through a LISA
Initiative (Laboratory for Interdisciplinary Advanced Simulation) 2012 grant [http://lisa.cilea.it],
and by a grant ``Dote ricerca'': FSE, Regione Lombardia.

\section{Appendix}
In Tables \ref{table:fitvar} and \ref{table:fitex} we give the fitting 
parameters of the energy as function of density with expressions \eqref{eosl}
and \eqref{eoss}. 
In our implementation of the QMC projection tecnhique, as $\psi_T$ we use
a \emph{Shadow Wave Function} (SWF).
Such a wave function, introduced by Vitiello \emph{et al.} \cite{swf1}, is known to provide a very accurate description
of the condensed phases of $^4$He \cite{swf3}: it has explicit pair correlations between the coordinates of the atoms as well as indirect many body correlations via some auxiliary \emph{shadow} variables, denoted  $\mathcal{S}=(\vec{s}_1,\dots,\vec{s}_N)$,
which are integrated over:

\begin{equation}
\label{swf}
\psi_T(\mathcal{R}) = \int \exp \left[ - \sum_{i<j}^N u_r(r_{ij}) - \sum_{i<j}^N v_s(s_{ij})  - \sum_{i}^N c \, |\vec{r}_i -\vec{s}_i |^2 \right] d\mathcal{S} 
\end{equation}
The pseudo-potentials are chosen to be a generalized McMillian form $u_r(r_{ij})=(b/r_{ij})^m$,
whereas the one for the shadow variables is chosen of the Aziz rescaled form
$v_s(s_{ij}) = \alpha v(\delta \, s_{ij})$. This SWF has the same form used by
Grisenti and Reatto \cite{gris} but as power $m$ we have used $m = 6$ because
this values improves the energy compared to $m = 5$ used in \cite{swf3}.
We have optimized the trial wave function \eqref{swf} varying the remaining
 variational parameters $b$, \,$\alpha$, \,$\delta$ and $c$ through Variational Monte Carlo simulations for various densities. Notice that the form of $\psi_T$ is the same for the liquid and for the solid, only the variational parameters take different values.
The optimized shadow wave functions is used as trial wave functions for
exact simulations at the same densities: the exact technique is
named Shadow Path Integral Ground State (SPIGS) method \cite{spigs1,spigs2}.

For reference purpose we give the optimal values of the SWF
variational parameters in tables \ref{table:spigsliq} and
\ref{table:spigssol} of the Appendix for the liquid and solid phases respectively.
The values of the exact and of the variational energy are also given
in that tables.
\begin{table}[t!]
\caption{\label{table:fitvar}
Values of the fit parameters for fitting functions \eqref{eosl} and \eqref{eoss} of the variational equation of state.}
\begin{ruledtabular}
\begin{tabular}{lrlr}
$E_0 \,(K)$     & $-0.753(3) $                          & $\alpha \, (K)$ &     $-15.2  \pm 28\%$ \\
$\rho_0 \, ($\AA$^{-2})$& $0.0393(2) $                  & $\beta \,  (K $\AA$^{2})$ &           $765    \pm 20\%$ \\
$A \, (K)$              & $1.39(6)    $                 & $\gamma \, (K $\AA$^{4})$ &   $-13311         \pm 13\%$ \\
$B \, (K)$              & $0.7(3)     $                 & $\delta \, (K $\AA$^{6})$ &   $80497  \pm 8.5\%$ \\
$C \, (K)$              & $0.93(15)   $                 &                           &                      \\ \hline
$\rho_f \, ($\AA$^{-2})$& $0.0677     $                 & $\rho_m \, ($\AA$^{-2})$  &   $0.0721$
\end{tabular}
\end{ruledtabular}
\end{table}

\begin{table}[t!]
\caption{\label{table:fitex}
Values of the fit parameters for fitting functions \eqref{eosl} and \eqref{eoss} of the exact equation of state.}
\begin{ruledtabular}
\begin{tabular}{lrlr}
$E_0 \,(K)$       & $-0.862(1)$           & $\alpha \, (a.u.)$ &  $-25.1  \pm 12\%$ \\
$\rho_0 \, ($\AA$^{-2})$& $0.0430(1)$             & $\beta \, (K $\AA$^{2})$ &            $1084   \pm 10\%$ \\
$A \, (K)$              & $2.00(3)  $             & $\gamma \, (K $\AA$^{4})$ &   $-16745         \pm 7.8\%$ \\
$B \, (K)$              & $2.1(1)   $             & $\delta \, (K $\AA$^{6})$ &   $92539  \pm 5.7\%$ \\
$C \, (K)$              & $0.52(14) $             &                           &                      \\ \hline
$\rho_f \, ($\AA$^{-2})$        & $0.0674$     & $\rho_m \, ($\AA$^{-2})$ &    $0.0701$
\end{tabular}
\end{ruledtabular}
\end{table}

\begin{table*}[t!]
\caption{\label{table:spigsliq}
Optimal values of the variational parameters for the SWFs at various densities in the liquid
phase, along with the values of the energy, computed both with variational and exact methods.
The results are compared with previous Green Function Monte-Carlo results $E_{GFMC}$ of
reference \cite{gfmc}.}
\begin{ruledtabular}
\begin{tabular}{cccccccc}
$\rho($\AA$^{-2})$ & $b($\AA$)$ & $c($\AA$^{-2})$ & $\delta (K) $ & $\alpha$ & $\hspace{2mm}E_{var} $ & $E_{SPIGS} $ & $E_{GFMC}(K)$ \\ \hline
$0.0310$ &        &         &         &         &                        &$-0.750(3)$                    &\\
$0.0321$ & $2.700$ & $0.710$ & $0.042$ & $0.918$ & $-0.710(1)$           &$-0.765(2)$                    &$-0.78(2)$ \\
$0.0332$ &        &         &         &         &                        &$-0.787(3)$                    &\\
$0.0358$ & $2.715$ & $0.720$ & $0.042$ & $0.920$ & $-0.739(2)$           &$-0.816(2)$                    &$-0.81(1)$ \\
$0.0421$ & $2.720$ & $0.710$ & $0.044$ & $0.920$ & $-0.747(2)$           &$-0.862(2)$                    &$-0.85(3)$ \\
$0.04315$ &         &         &         &         &                      &$-0.861(2)$                    &\\
$0.0478$ & $2.720$ & $0.700$ & $0.044$ & $0.920$ & $-0.675(2)$           &$-0.835(3)$                    &\\
$0.0490$ &         &         &         &         &                       &$-0.817(2)$                    &\\
$0.0536$ & $2.720$ & $0.700$ & $0.043$ & $0.900$ & $-0.516(3)$           &$-0.704(2)$                    &$-0.67(3)$ \\
$0.0600$ & $2.715$ & $0.660$ & $0.042$ & $0.870$ & $-0.189(3)$           &$-0.404(2)$                    & \\
$0.0658$ & $2.710$ & $0.645$ & $0.042$ & $0.840$ & $\hspace{3mm}0.295(2)$       &$\hspace{3mm}0.065(3)$ & $-0.01(4)$\\
$0.0719$ & $2.710$ & $0.635$ & $0.042$ & $0.820$ & $\hspace{3mm}1.057(3)$       &$\hspace{3mm}0.798(3)$ & $\hspace{3mm}0.82(4)$\\
\end{tabular}
\end{ruledtabular}
\end{table*}

\begin{table*}[t!]
\caption{\label{table:spigssol}
Optimal values of the variational parameters for the SWFs at various densities in the solid phase,
along with the values of the energy, computed both with variational and exact methods.
The results are compared with previous Green Function Monte-Carlo results $E_{GFMC}$ of
reference \cite{gfmc}; values with $\star$ are computed near the given density.}
\begin{ruledtabular}
\begin{tabular}{cccccccc}
$\rho$(\AA$^{-2})$ &     $b ($\AA$)$    &        $c ($\AA$^{-2})$       &       $\delta$(K)     &       $\alpha$ & $\hspace{2mm}E_{var}$ & $E_{SPIGS}$ & $E_{GFMC}(K)$ \\ \hline
$0.0690$ &         &         &         &         &      & $ 0.441(5)$ &         \\
$0.0740$ & $2.705$ & $0.470$ & $0.183$ & $0.810$ & $1.186(2)$ & $0.951(2)$ &            \\
$0.0765$ & $2.710$ & $0.500$ & $0.184$ & $0.835$ & $1.504(3)$ & $1.292(3)$ & $1.30(2)$  \\
$0.0835$ & $2.700$ & $0.650$ & $0.182$ & $0.860$ & $2.789(3)$ & $2.579(2)$ & $2.78(7)^\star$ \\
$0.0905$ & $2.710$ & $0.700$ & $0.183$ & $0.880$ & $4.733(2)$ & $4.483(2)$ & $4.91(3)^\star$ \\
$0.0975$ & $2.705$ & $0.800$ & $0.182$ & $0.900$ & $7.520(3)$ & $7.223(3)$ &    \\
\end{tabular}
\end{ruledtabular}
\end{table*}


\begin{thebibliography}{99}
\bibitem{az79} R.A. Aziz, V.P.S. Nain, J.S. Carley, W.L. Taylor and G.T. McConville,
               {\em J. Chem. Phys} {\bf 70}, 4330 (1979).
\bibitem{legg} A.J. Leggett, {\em Rev. Mod. Phys.} {\bf 71}, S318 (1999).
\bibitem{land} L.D. Landau, {\em J. Phys. USSR} {\bf 5}, 71 (1941); {\em J. Phys. USSR} {\bf 11}, 91 (1947).
\bibitem{fey1} R.P. Feynman, {\em Phys. Rev.} {\bf 90}, 1116 (1953).
\bibitem{cowo} D. G. Henshaw and A.D.B. Woods, {\em Phys. Rev.} {\bf 121}, 1266 (1961).
\bibitem{fey2} M. Cohen and R.P. Feynman, {\em Phys. Rev.} {\bf 120}, 1189 (1956).
\bibitem{cbf} E. Manousakis and V.R. Pandharipande {\em Phys. Rev. B} {\bf 30}, 5062 (1984).
\bibitem{eswf} W. Wu, S.A. Vitiello, L. Reatto, and M.H. Kalos. {\em Phys. Rev. Lett.} {\bf 67}, 1446 (1991).
\bibitem{swf2} D.E. Galli, E. Cecchetti, and L. Reatto, {\em Phys. Rev. Lett.} {\bf 77}, 5401 ( 1996).
\bibitem{roto} L. Reatto and D.E. Galli, {\em  Int. J. Mod. Phys. B} {\bf 13}, 607 (1999).
\bibitem{feyl} R.P. Feynman, {\em Statistical Mechanics} (W.A. Benjamin Inc., New York, 1972).
\bibitem{gube} R.N. Silver, D.S. Sivia, and J.E. Gubernatis, {\em Phys. Rev. B} {\bf 41}, 2380 (1990).
\bibitem{cepe} M. Boninsegni, and D.M. Ceperley, {\em J. Low Temp. Phys.} {\bf 104}, 339 (1996).
\bibitem{rept} S. Baroni, and S. Moroni, {\em Phys. Rev. Lett.} {\bf 82}, 4745 (1999).
\bibitem{gift} E. Vitali, M. Rossi, L. Reatto and D. E.Galli, {\em Phys. Rev. B}, {\bf 82}, 174510 (2010).
\bibitem{asm} A.W. Sandvik, {\em Phys. Rev. B} {\bf 57}, 10287 (1998); O.F. Sylju{\aa}sen, {\em Phys. Rev. B} {\bf 78}, 174429 (2008).
\bibitem{gift2} M. Rossi, E. Vitali, L. Reatto, D.E. Galli, {\em Phys. Rev. B} {\bf 85}, 014525 (2012).
\bibitem{gift3} M. Nava, D.E. Galli, M.W. Cole, L. Reatto, {\em J. Low Temp. Phys.}, DOI 10.1007/s10909-012-0770-9 (2012).
\bibitem{ktb} J.M. Kosterlitz and D. J. Thouless, {\em J. Phys. C: Solid State Phys} {\bf 5}, 124 (1972).
\bibitem{reppy} D. J. Bishop, and J. D. Reppy, {\em Phys. Rev. Lett.} {\bf 40}, 1727 (1978).
\bibitem{gfmc} P.A. Whitlock. G.V. Chester, and M.H. Kalos. {\em Phys. Rev. B} {\bf 38}, 2418 (1988).
\bibitem{exp2} B.E. Clernents, E. Krotscheck, and C.J. Tymczak. {\em Phys. Rev. B} {\bf 53} 12253 (1996).
\bibitem{godfrin}  H. J. Lauter, H. Godfrin, P. Leiderer, {\em J. Low Temp. Phys.} {\bf 87}, 425 (1992).
\bibitem{cole} M. W. Cole, D. R. Frankl, and D. L. Goodstein, {\em Rev. Mod. Phys.} {\bf 53} 199 (1981).
\bibitem{cleme} B. E. Clements, H. Forbert, E. Krotscheck, H. J. Lauter, M. Saarela, and C. J. Tymczak,
                {\em Phys. Rev. B} {\bf 50}, 6958 (1994).
\bibitem{gris} R.E. Grisenti, and L. Reatto, {\em J. Low Temp. Phys.}, {\bf 109} (1997).
\bibitem{pigs} A. Sarsa, K.E. Schmidt and W.R. Magro, {\em J. Chem. Phys.} {\bf 113}, 1366 (2000).
\bibitem{spigs1} D.E. Galli and L. Reatto, {\em Mol. Phys.} {\bf 101}, 1697 (2003).
\bibitem{spigs2} D.E. Galli, and L. Reatto, {\em J. Low Temp. Phys.} {\bf 136}, 343 (2004).
\bibitem{gift4} M. Nava, A. Motta, D.E. Galli, E. Vitali, S. Moroni, {\em Phys. Rev. B} {\bf 85}, 184401 (2012).
\bibitem{patate} M. Rossi, M. Nava, L. Reatto, and D.E. Galli, {\em J. Chem. Phys.} {\bf 131}, 154108 (2009).
\bibitem{pair} D.M. Ceperley, {\em Rev. Mod. Phys.} {\bf 67}, 279 (1995).
\bibitem{rech} L. Reatto and G. V. Chester, {\em Phys. Rev.} {\bf 155} 88 (1967).
\bibitem{cowo2} R.A. Cowley, and A.D.B. Woods, {\em Can. J. Phys.} {\bf  49}, 177 (1971);
               A.D.B. Woods and R.A. Cowley, {\em Rep. Prog. Phys.} {\bf 36}, 1135 (1973).
\bibitem{pori} T.R. Prisk, N.C. Das, S.O. Diallo, G. Ehlers, A.A. Podlesnyak, N. Wada, S. Inagaki, P.E. Sokol, arXiv:1211.0350. 
\bibitem{swf1} S.A. Vitiello, K.J. Runge and M.H. Kalos, {\em Phys. Rev. Lett.} {\bf 60}, 1970 (1988).
\bibitem{swf3} B. Krishnamachari and G.V. Chester, {\em Phys. Rev. B} {\bf 61}, 9677 (2000).
\end{thebibliography}
\end{document}